\documentclass[draftcls,12pt,onecolumn]{IEEEtran}

\usepackage{cite}
\usepackage[pdftex]{graphicx}
\usepackage[cmex10]{amsmath}
\usepackage{amssymb}
\usepackage{amsfonts}

\usepackage{algorithm}
\usepackage{algorithmic}
\usepackage[caption=false,font=footnotesize]{subfig}
\usepackage{color}
\usepackage{array}
\usepackage{mdwmath}
\usepackage{mdwtab}
\usepackage{eqparbox}
\usepackage[colorlinks=true,urlcolor=black]{hyperref}
\usepackage{epstopdf}
\usepackage{cite}
\usepackage{amsmath,amssymb,amsfonts}
\usepackage{algorithmic}
\usepackage[pdftex]{graphicx}
\usepackage{textcomp}
\usepackage{epstopdf}
\usepackage{xcolor}
\newtheorem{remark}{Remark}

\usepackage{multirow}
\usepackage{bigstrut}
\usepackage{booktabs}
\usepackage{makecell}
\usepackage{color}
\usepackage{algorithm}
\usepackage{algorithmic}
\usepackage[caption=false,font=footnotesize]{subfig}
\usepackage{color}

\begin{document}
\title{3D Spectrum Mapping Based on ROI-Driven UAV Deployment}
\author{ \IEEEauthorblockN{Qihui Wu, \emph{Senior Member, IEEE}, Feng Shen,\\ Zheng Wang, \emph{Member, IEEE}, and Guoru~Ding, \emph{Senior Member, IEEE}}

\thanks{Qihui Wu, Feng Shen (corresponding author) and Zheng Wang are with Key Laboratory of Dynamic Cognitive System of Electro-magnetic Spectrum Space (Nanjing Univ. Aeronaut. Astronaut.), Ministry of Industry and Information Technology; Guoru Ding is with Army Engineering University of PLA.}
}
\date{}
\maketitle

\begin{abstract}
 Given the explosive growth of Internet of Things (IoT) devices ranging from the two-dimensional (2D) ground to the three-dimensional (3D) space, it is a necessity to establish a 3D spectrum map to comprehensively present and effectively manage the 3D spatial spectrum resources in smart city infrastructures. By leveraging the popularity and location flexibility of the unmanned aerial vehicles (UAVs), we are able to execute spatial sampling with these emerging flying spectrum-monitoring devices (SMDs) at will.
In this paper, we first present a brief survey to show the state-of-the-art studies on spectrum mapping. Then, we introduce the 3D spectrum mapping model.
Next, we propose a 3D spectrum mapping framework which is composed of pre-sampling, spectrum situation estimation, UAV deployment and spectrum recovery. Therein we develop a Region of Interest (ROI)-driven UAV deployment scheme, which selects new sampling points of the highest estimated interest and the lowest energy cost iteratively. Meanwhile, we slice the entire 3D spectrum map into a series of ``images" and ``repair" those unsampled locations. Furthermore, we provide an exemplary case study on the 3D spectrum mapping, where, for example, an important event is being held and the entire spectrum situation needs to be monitored in real time to deal with malicious interference sources. Lastly, the challenges and open issues are discussed.
\end{abstract}


\IEEEpeerreviewmaketitle


\section{INTRODUCTION}
With the rapidly increasing number of wireless devices and the scarcity of spectrum resources, how to efficiently utilize spectrum resources and manage the dramatic data produced by mobile internet and Internet of Things (IoT) has received much attention \cite{IoT-survey}.
As reported by the Federal Communications Commission (FCC) \cite{fcc}, some unlicensed bands such as industrial, scientific, medical bands and licensed bands for land mobile communications are overcrowded. However, most of licensed bands allocated to for example broadcasting TVs or analogue cellular telephony are often used inefficiently. Therefore, being able to ``see" the spectrum in the wireless environment offers high value for wise operations of IoT devices, which is called spectrum situational awareness in the DARPA RadioMap program \cite{radiomap}.
Spectrum situation refers to the current state of the electromagnetic environment, including spectrum busy or idle state, spectrum signal strength, spectrum modulation mode, and spectrum access protocol, etc. The main goal of spectrum situational awareness is to obtain the current state of the spectrum space which is the cornerstone of spectrum state generation and utilization. If we correspond the spectrum situational awareness results to the 3D geographic locations one by one, we can construct a map describing the signal strength at any location in space, and the process of obtaining this map is called spectrum mapping.

Spectrum devices in IoT can be grouped into two classes: spectrum-monitoring devices (SMDs) and spectrum-utilizing devices (SUDs). SMDs are responsible for monitoring or sensing spectrum signal strength, whereas SUDs utilize the spectrum to transmit data \cite{futureiot}.
If the SUDs are offered a 3D spatial spectrum map, they can immediately access the idle spectrum and avoid conflicts where it is busy. They can also help to update the spectrum map by sharing their spectrum situational awareness results. This increases the number of devices and the amount of communications that can be supported in limited spectrum environment. Thus, the spectrum utilization will be undoubtedly improved.

Recently, the increasing popularity of UAVs in both civil and military fields has been witnessed \cite{uavpopularity}, which on one hand, makes UAVs a new part of the SUDs in 3D spectrum space, while on the other hand, poses promising potentials to explore and construct a 3D spectrum map by properly leveraging the location flexibility of flying UAV SMDs.
What's more, due to the heterogeneity of the 3D electromagnetic environment, regions of environment are allowed to have different interests or priority levels. With this concept, the 3D target area can be divided into small cubes for measuring, and each cube has an importance value. Unlike traditional sampling which assumes that all locations have the same priority, positions with large values of interest are supposed to be sampled first \cite{PrioritizedSensing}.

In this article, we firstly present a brief survey on the state-of-the-art studies on spectrum mapping. Then, we describe the 3D spectrum mapping model. Next, we propose a 3D spectrum mapping framework which consists of four key components: pre-sampling, spectrum situation estimation, iterative ROI-driven UAV deployment and spectrum map recovery.
Furthermore, we provide an exemplary case study on the performance gain of the proposed spectrum mapping framework. Finally, the challenges and open issues are given, followed by the conclusion of our work.

\section{SPECTRUM MAPPING: STATE-OF-THE-ART}
At present, there are a few studies on spectrum mapping.
In \cite{rem}, the authors propose a prototype of a radio environment map (REM) which aims at storing and reasoning the spectrum data obtained from heterogeneous spectrum sensors. This is a basic guiding work of constructing a radio environment map for spectrum management.

In \cite{Dingkernel2013}, the authors apply kernel-based learning to cognitive radio networks. In the third application of the paper, they use nonlinear Support Vector Machine (SVM) to detect the boundary between the coverage area and the non-coverage area of the primary user. This is actually a prototype of a 2D binary spectrum map.
However, it distinguishes the coverage area and the non-coverage area of one signal source. This is obviously an idealization of the actual cognitive radio network. The actual spectrum map should be three-dimensional, and it should be spatial-heterogeneous, which means the spectrum mapping in height dimension can not be ignored as discussed in \cite{uav3dsensing2019}.

In \cite{sun2018}, the authors formulate the spectrum situation of multiple frequency points with multiple time slots as an ``image" and propose an idea of image inference to complete and predict the spectrum data. However, in these works, the spectrum data is only in time-frequency dimension, and the spectrum situation is not combined with the geographical location, failing to present a 3D spectrum map.

Moreover, \cite{PrioritizedSensing} proposes the first priority-estimation method for area-priority-based wireless sensor networks deployments, which uses K-means clustering information to determine the priorities of the sensor coverage areas. However, it is aimed at 2D scenarios, and the goal is to maximize the coverage with position-fixed sensors, which is not suitable for flexible UAV scenarios.

In a nutshell, it is observed that although the topic of spectrum mapping has been studied, most of existing studies focus on 2D ground spectrum mapping and analyze the binary map of spectrum occupy states with position-fixed sensors, which doesn't match the practical 3D spectrum environment. There are no reports on 3D spectrum mapping with channel measured values in spectrum-heterogeneous environment by leveraging flying UAV SMDs.
\begin{figure*}[!t]
\centering
\includegraphics[width=.9\linewidth]{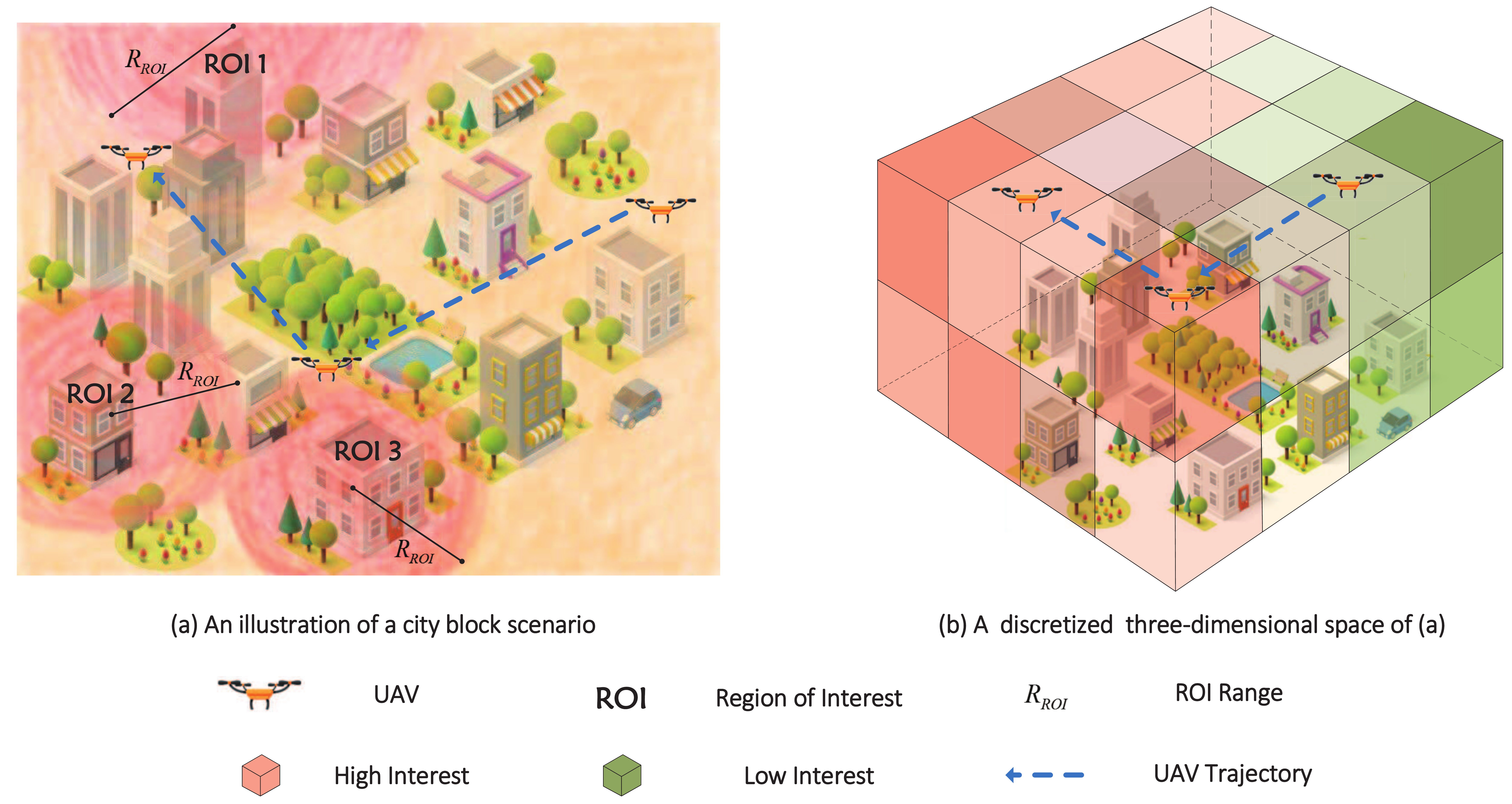}
\caption{System Model. (a) is an illustration of a city block, there are several Regions of Interest (ROIs) in the scenario and ${R_{ROI}}$ represents the radius of ROI. Every position within ROI is Point of Interest (POI) that we expect to sample with UAVs. (b) is a 3D tensor obtained by discretizing the 3D space in (a). Each cube has an interest value, the red means high interest, and the green represents low interest.}
\label{scenario1}
\end{figure*}

\section{3D SPECTRUM MAPPING MODEL}
We consider constructing a spectrum map of a city block scenario as shown in Fig. \ref{scenario1}, in which spectrum from building facilities, vehicles, human being, etc. spreads everywhere.
As illustrated in Fig. \ref{scenario1} (a), there are some areas in red that we need to pay special attention than the other areas. We call these areas as the Regions of Interest (ROIs) and the point in the ROI as the Point of Interest (POI).
There often exists sources in the ROIs, so we take the source as the center and define the ROI size with a 3D spherical space of radius ${R_{ROI}}$.

In order to solve the task of 3D spectrum mapping, we discretize the entire 3D space into small cubes as shown in Fig. \ref{scenario1} (b). The color of each cube represents the degree of interest of this position. Red indicates high degree of interest and green indicates low interest. This constitutes a spectrum tensor $\mathcal{X}  \in {\mathbb{R}^{{N_1} \times {N_2} \times {N_3}}}$ of the 3D space, where ${N_1}$, ${N_2}$, and ${N_3}$ represent the grids number of tensor $\mathcal{X}$ in $x$, $y$, $z$ dimensions, respectively. We define $x = 1,2,...,{N_1},y = 1,2,...,{N_2},z = 1,2,...,{N_3}$, and $(x,y,z)$ denotes the $(x,y,z)$-th small cube in the tensor $\mathcal{X}$.

In the process of spectrum mapping, the UAVs are dispatched to perform spatial spectrum signal strength measurements in every cube.
For an unknown 3D space, we pay more attention to the mapping accuracy of the regions of interest. For the uninterested regions, even if the mapping error is large, it will not affect much. Therefore, our goal is to operate spectrum sensing in as many ROIs as possible by leveraging UAV's movement flexibility and recover the energy values of all unsampled cubes to get the complete 3D spectrum map.
If we sample every location, we will undoubtedly obtain the accurate spectrum map, however, it will suffer enormous resource consumption.
Therefore, we need to let the UAV adaptively search for suitable sampling positions to avoid blind sampling. Through partial sampling, we can exploit the spatial correlation among the sampling points to restore the spectrum situation of all ROIs with minimized restored error.
Of course, if the sampling ratio is too small, the spatial correlation among all the sampling points will be severely damaged, resulting in poor recovery accuracy. In addition, even with the same sampling ratio and sampling location, different UAV trajectories will lead to different energy consumptions.

Thus, the objective is to spend less energy and simultaneously obtain better ROI spectrum situation recovery, subject to limited sampling ratio by optimizing the sequence of sampling points as follows:
\begin{equation}
{S^*} = \mathop {\rm{arg \\\ min }}\limits_S {\rm{ }}\xi {\rm{(}}\alpha {W_{ROI}})\gamma (\beta E),\label{eq:minwe}
\end{equation}
\begin{align}
&~\text{\emph{s.t.}}~r = \frac{{{N_s}}}{{{N_1} \times {N_2} \times {N_3}}},\tag{C\ref{eq:minwe}-1}\\
&~~~~~{W_{ROI}} = \frac{1}{{{N_{ROI}}}}{\sum\limits_{i \in ROI} {\left( {\frac{{\left| {{{\cal T}_{{\cal X}_R^i}} - {{\cal T}_{{{\cal X}^i}}}} \right|}}{{{{\cal T}_{{{\cal X}^i}}}}}} \right)} ^2},\tag{C\ref{eq:minwe}-2}\\
\notag
\end{align}
where $S$ represents a sequence of sampling points, including the number of samples and the sampling positions. $W_{ROI}$ is the Relative Mean Square Error ($RMSE$) of the ROIs. $E$ is the UAV consumption for obtaining the samples in $S$. ${N_s}$ denotes the sampling number, and $r$ is the sampling ratio. $\alpha$ and $\beta$ represent the positive weight parameters which can be adjusted according to importance assigned between recovery error $W_{ROI}$ and energy consumption $E$. $\xi ( \cdot )$ and $\gamma ( \cdot )$ are correction functions of $W_{ROI}$ and $E$ to avoid the situation that $W_{ROI}$ and $E$ are rapidly approaching zero when the sampling ratio is too large and too small, respectively. In (C1-2), the ${N_{ROI}}$ represents the total points number in all ROIs. ${{{\cal T}_{{{\cal X}^i}}}}$ and ${{{\cal T}_{{\cal X}_R^i}}}$ represent the original and recovered spectral energy value at the $i$-th position, respectively. Here, the signal energy strength ${{{\cal T}_{{{\cal X}^i}}}}$  for each location is generated by linear superposition of signals from all sources after path loss and is then added with Gaussian additive white noise.

Our goal in Eq. (1) is a function of $S$, which means given a sampling ratio $r$, we can find a total of $\frac{{({N_1} \cdot {N_2} \cdot {N_3})!}}{{((1 - r) \cdot {N_1} \cdot {N_2} \cdot {N_3})!}}$ deployment solutions. This scale is too large to be solved by means of exhaustive search.
With the assigned different importance values and priorities, we pay more attention to the recovery accuracy of the ROIs and our objective is to make the recovery error of the ROIs as small as possible. Thus, it is beneficial to sample points with large POI values and the problem can be turned into finding a $S$ of the largest POI value accumulation with the minimum energy consumption, namely:
${S^*}{\rm{ = }}\mathop {{\rm{arg \\ \  max}}}\limits_S {\rm{ }}\frac{{\xi \left( {\alpha \sum\limits_{i \in S} {{\cal P}{\cal O}{{\cal I}_{{{\cal X}^i}}}} } \right)}}{{\gamma \left( {\beta {E_S}} \right)}}$,
where ${\cal P}{\cal O}{{\cal I}_{{{\cal X}^i}}}$ represents the POI value of the sampling point ${{\cal X}^i}$. Here it is defined as the
spectral energy value, the higher the energy value, the higher the POI value.
This problem is similar but different to the traveling salesman problem (TSP) \cite{tsp}.
\begin{remark}\emph{In the TSP problem, you need to find the shortest path connecting all the given target positions. However, in this problem, we must not only determine which target locations to choose, but also make the path connecting these locations the shortest given the sampling rate.}
\end{remark}

Since we have no priori knowledge of the target 3D space before dispatching the UAVs, a simple UAV deployment method is random deployment. That is, the ${N_1} \cdot {N_2} \cdot {N_3}$ points in the 3D space are selected with the same probability, which also accords with the maximum information entropy principle.
Although the random deployment scheme is easy to operate, there will still be a large number of sampling points out of the ROIs, resulting in a huge recovery error for our target ROIs.
Therefore, we propose an \emph{iterative} 3D spectrum mapping framework based on ROI coping with the difficulty in accurate spectrum mapping without priors.

\begin{figure*}[!t]
\centering
\includegraphics[width=.9\linewidth]{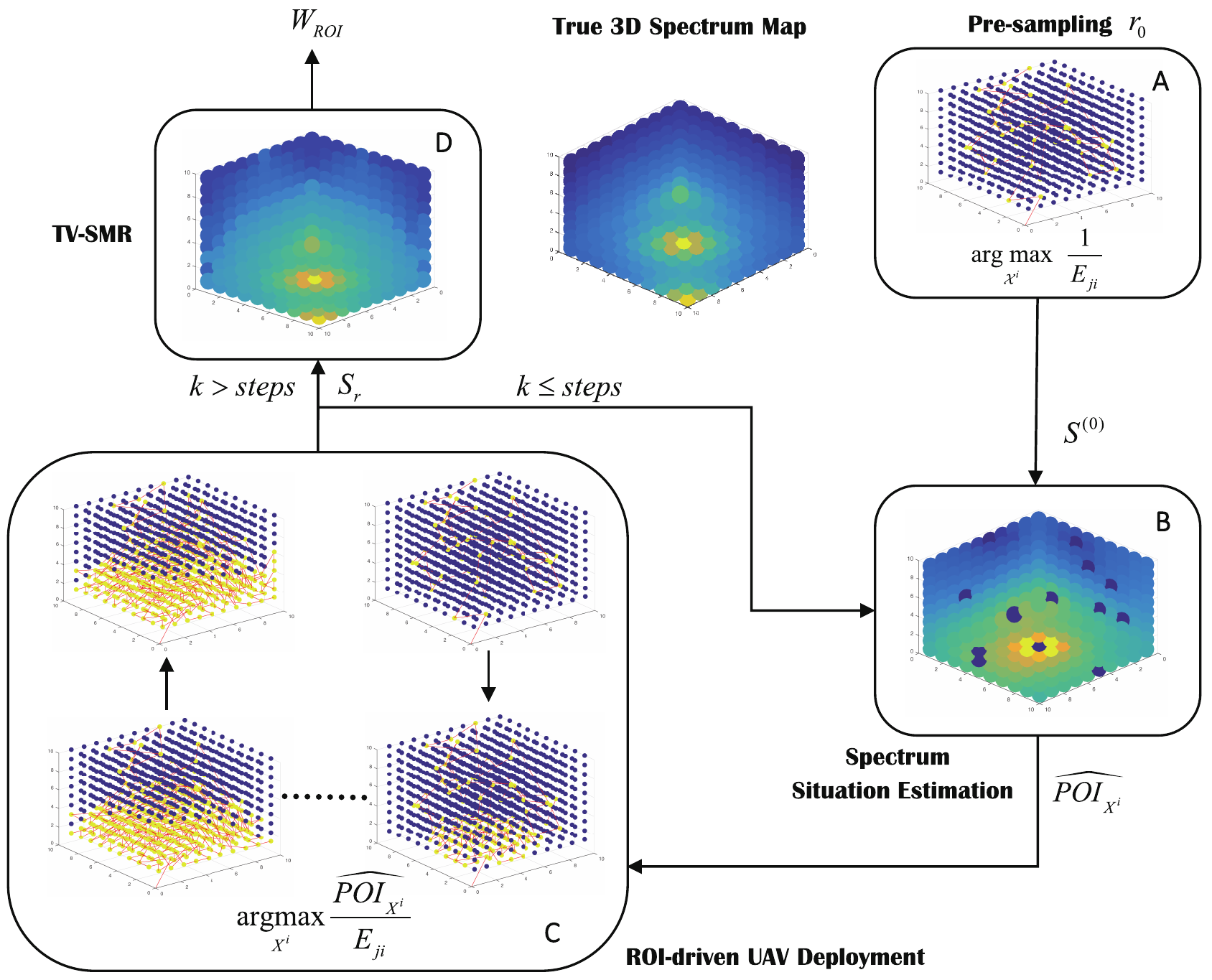}
\caption{Graphical illustration of 3D spectrum mapping framework. The whole framework consists of four parts: pre-sampling, spectrum situation estimation, ROI-driven UAV deployment and spectrum map recovery. On the right, the UAV first obtains some samples of ratio $r_0$ as prior information and gets the initial random sampling sequence ${S^{(0)}}$. Then, we execute spectrum situation estimation and obtain the estimated interest $\widehat {{\cal P}{\cal O}{{\cal I}_{{{\cal X}^i}}}}$ on each unsampled point. Afterwards, according to Eq. (3), we operate ROI-driven UAV deployment and get the newly-added samples of ratio $r_p$ for ${r_p} \cdot {N_1} \cdot {N_2} \cdot {N_3}$ times in the $k$-th step. If $k \le steps$ $(steps=(r-r_0)/r_p)$, then we run the spectrum situation estimation and ROI-driven UAV deployment again. When $k > steps$, we get all the samples $S_r$ corresponding to sampling ratio $r$. Finally, we execute the TV-SMR algorithm and work out the RMSE $W_{ROI}$ with the true 3D spectrum map.   }
\label{framework}
\end{figure*}

\section{3D SPECTRUM MAPPING FRAMEWORK}
This section will introduce our proposed ROI-driven 3D spectrum mapping framework. As shown in Fig. \ref{framework}, the entire framework includes four parts: pre-sampling, spectrum situation estimation, UAV deployment, and spectrum map recovery. The details are as follows:
\subsection{Pre-sampling}
Because we do not have any prior information on the target area, we can first randomly obtain some samples by pre-sampling, and then select the new sampling points and UAV trajectories by analyzing the obtained samples. For example, if the sampling ratio is $r$, we can randomly select $r_0$ percent of all the positions as pre-sampling points. Here we consider the random deployment scheme in the pre-sampling step, that is: ${{\cal X}^{i*}}{\rm{ = }}\mathop {{\rm{arg max}}}\limits_{{{\cal X}^i}} {\rm{ }}\frac{1}{{{E_{ji}}}}$,
where ${{{E_{ji}}}}$ represents the total energy consumption for flying from ${{{\cal X}^j}}$ to ${{{\cal X}^i}}$ and then hovering to sample at ${{{\cal X}^i}}$. Then, we obtain the initial sampling sequence ${S^{(0)}}$ with the least energy consumption. We denote all sampled points as $S$, so at the first step of spectrum mapping $S = {S^{(0)}}$.
\subsection{Spectrum Situation Estimation}
This section mainly describes how to exploit the already sampled points to estimate the situation of the entire 3D spectrum space, that is, to estimate the POI values of all unsampled points. Here we use the Inverse Distance Weight (IDW) method \cite{idw} which is also called the Distance Reciprocal Multiplication method.
It assumes closer things are more similar than things that are far away. When predicting any unmeasured position, the IDW method uses the measured values around the predicted position. Compared with the measurement value that is far from the predicted position, the measurement value that is closer to the predicted position has a greater influence on the predicted value. The IDW method assumes that each measurement point has a local effect, and this effect decreases with increasing distance. Thus, we have our spectrum situation estimation as follows:
\begin{equation}
\widehat {{\cal P}{\cal O}{{\cal I}_{{{\cal X}^i}}}} = \sum\limits_{j \in S} {{\omega _{{{\cal X}^j}}}{\cal P}{\cal O}{{\cal I}_{{{\cal X}^j}}}} ,{{\cal X}^i} \in \bar S,
\end{equation}
where $\widehat {{\cal P}{\cal O}{{\cal I}_{{{\cal X}^i}}}}$ is the estimated POI value of ${{{\cal X}^i}}$, ${\omega _{{{\cal X}^j}}} = 1/(d_{ij}^n\sum\limits_{j \in S} {\frac{1}{{d_{ij}^n}}} )$, $\bar S$ represents the complementary set of $S$. ${{\omega _{{{\cal X}^j}}}}$ denotes the normalized influence weight of the point ${{{\cal X}^j}}$ on ${{{\cal X}^i}}$. $n$ is a positive adjustable parameter and usually $n=2$.
\subsection{ROI-Driven UAV Deployment}
After the spectrum situation estimation process, we need to select new points. Here, we not only consider the estimated POI value, but also the energy consumption required for the new samples selection.
\subsubsection{Energy Consumption}
The UAV flies through points to obtain the signal energy value. If the flight distance is too long, the energy consumption of spectrum mapping will be very large. During the flight, there are four states including horizontal flying, vertical upward flying, vertical downward flying, and hovering. The energy consumption formulas for these four states are studied in \cite{energymodel}. Thus, the energy consumption of a rotor UAV from one sampling position to the next sampling position can be divided into flying energy consumption ${E_{route}}$ and hovering energy consumption ${E_{hover}}$. ${E_{hover}}$ is proportional to the hovering time. ${E_{route}}$ depends on the relative position between ${{\cal X}^j}$ and ${{\cal X}^i}$, which may be on the same horizontal plane or on the same vertical plane. But in most cases, it is oblique flying between ${{\cal X}^j}$ and ${{\cal X}^i}$.
For the oblique flying, we define the energy consumption as:
${E_{route}} = (1 - \cos \theta  )E_{ji}^{vertical} + \cos \theta E_{ji}^{horizonal}$, where $E_{ji}^{horizonal}$ and $E_{ji}^{vertical}$ are the energy consumptions in horizonal and vertical directions, respectively. $\theta $ is the absolute value of elevation angle between ${{\cal X}^j}$ and ${{\cal X}^i}$. $\theta  = 0$ and $\frac{\pi }{2}$ correspond to horizontal flying and vertical flying, respectively.
\subsubsection{New Sample Selection}
Based on the above estimated POI values, UAV energy consumption analysis, and the optimization problem, we get a suboptimal solution, that is, iteratively selecting new sampling points with the largest estimated POI and the lowest consumption. The expression is:
\begin{equation}
{{\cal X}^{i*}}{\rm{ = }}\mathop {{\rm{arg max}}}\limits_{{{\cal X}^i}} {\rm{ }}\frac {\widehat {{\cal P}{\cal O}{{\cal I}_{{{\cal X}^i}}}}} {{{E_{ji}}}},
\end{equation}
where $\widehat {{\cal P}{\cal O}{{\cal I}_{{{\cal X}^i}}}}$ represents the estimated POI value of ${{{\cal X}^i}}$. ${{E_{ji}}}$ represents the total energy consumption for flying from ${{{\cal X}^j}}$ to ${{{\cal X}^i}}$ and then hovering to sample at ${{{\cal X}^i}}$.
When the ratio of newly-added samples is $r_p$, we repeat the operation of formula (3) for ${r_p} \cdot {N_1} \cdot {N_2} \cdot {N_3}$ times, and denote the total newly-added sampling points per step as $\Delta S$.
Then, the total sampling points for the $(k+1)$-th spectrum situation estimation is ${S^{(k)}} = {S^{(k-1)}} + \Delta {S^{(k-1)}}$ $(k \ge 1)$.
According to Eq. (3) the objective function aims to generate UAV trajectory point-by-point, which can effectively select key positions and generate the suboptimal trajectory. Finally, we will obtain all samples ${S_r} = {S^{(steps )}}$ of sampling ratio $r$ where $steps = (r - {r_0})/{r_p}$.
\begin{figure*}[!t]
\centering
\includegraphics[width=.9\linewidth]{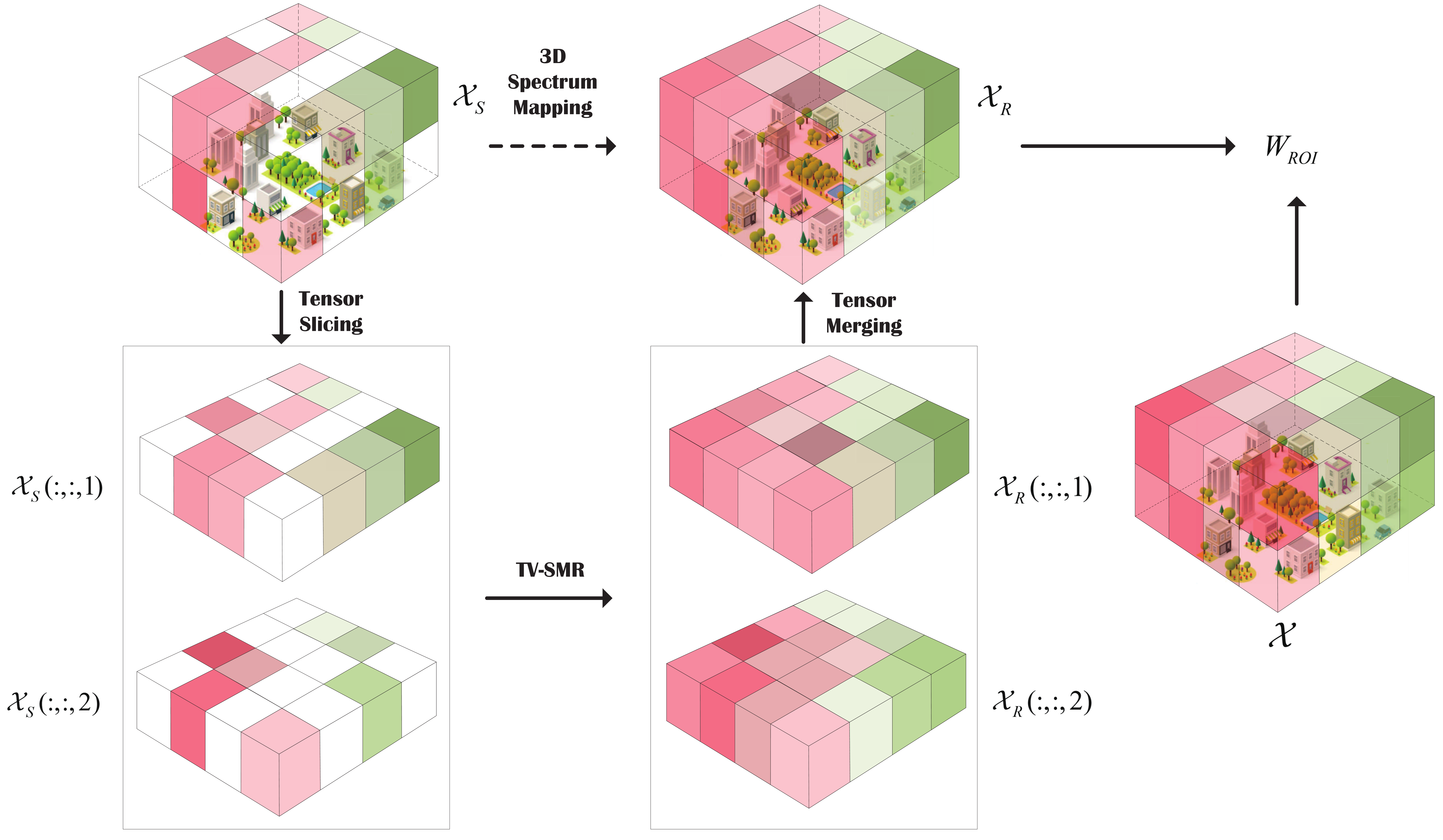}
\caption{ Graphical illustration of total variance-based 3D spectrum map recovery (TV-SMR) algorithm. We slice the 3D sparse spectrum map into ${{\cal X}_S}(:,:,1)$ and ${\mathcal{X}_S}(:,:,2)$. Then we recover each slice and get ${{\cal X}_R}(:,:,1)$, ${\mathcal{X}_R}(:,:,2)$. Finally, we merge all recovered slices and get the recovered spectrum map ${\mathcal{X}_R}$ which is compared with the real spectrum map to get the error $W_{ROI}$.}
\label{tvsmr}
\end{figure*}

\begin{figure*}[!t]
\centering
\includegraphics[width=1.05\linewidth]{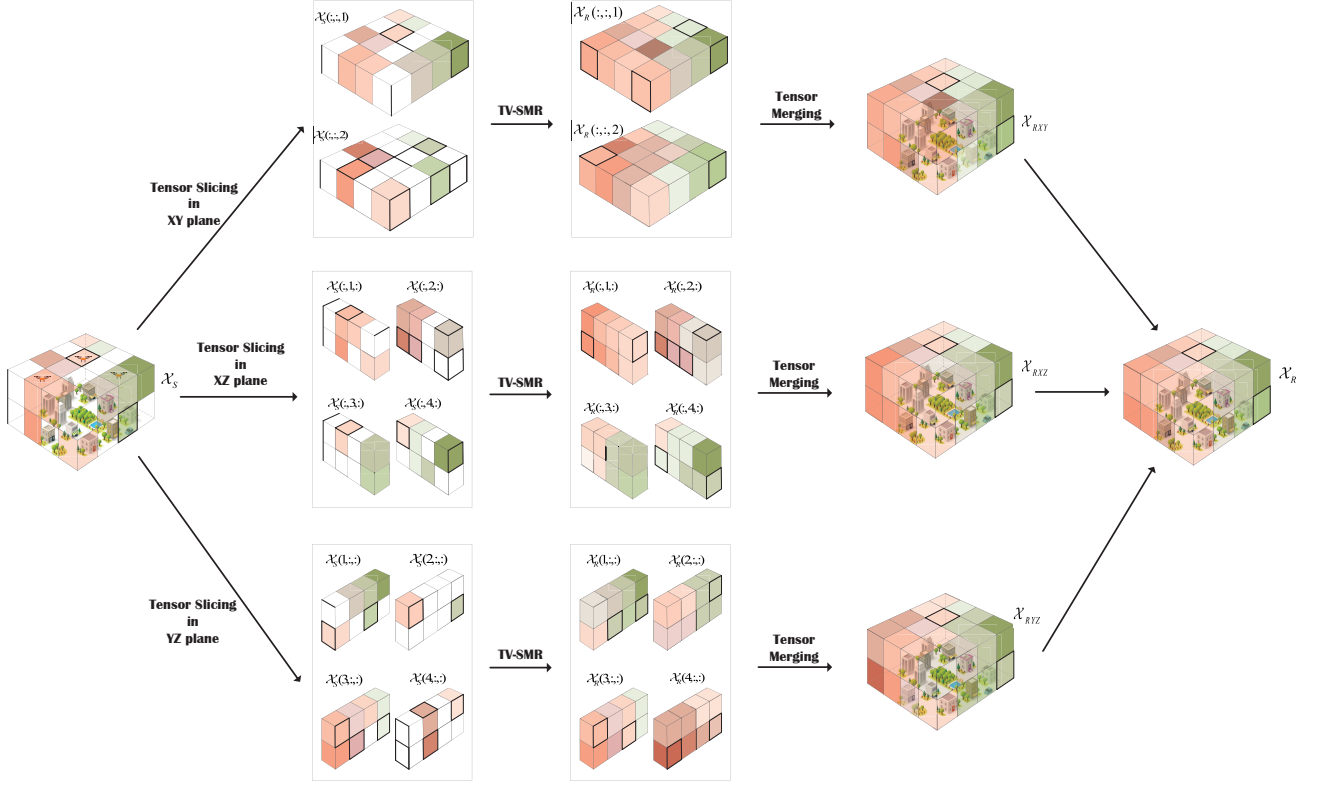}
\caption{3DTV-SMR algorithm is an improved TV-SMR algorithm. We execute the TV-SMR algorithm in three different directions, $XY$ plane, $YZ$ plane, and $ZX$ plane, and finally fuse the three recovery results together. This can exploit the spatial relations in different directions and avoid the loss caused by unidirectional slicing.}
\label{3dtvsmr}
\end{figure*}
\subsection{Total Variance-based Spectrum Map Recovery}
After we get all samples ${S_r}$ of a set sampling ratio $r$, we need to mine the relationship among all sampling cubes. We innovatively turn the problem of missing value completion for the 3D spectrum map into a problem of damaged-image restoration. As shown in Fig. \ref{tvsmr}, the ${S_r}$ forms a sampling tensor $\mathcal{X}_S$. Then, we perform the tensor slicing operation on the 3D sparse spectrum map tensor $\mathcal{X}_S$, and obtain two images ${{\cal X}_S}(:,:,1)$ and ${\mathcal{X}_S}(:,:,2)$. Of course, for $\mathcal{X}  \in {\mathbb{R}^{{N_1} \times {N_2} \times {N_3}}}$, we will get ${N_3}$ slices, ${{\cal X}_S}(:,:,1)$, ${\mathcal{X}_S}(:,:,2)$,..., ${\mathcal{X}_S}(:,:,N_3)$, each slice is a ${N_1} \times {N_2}$ matrix.
After we recover all the slices and merge them, we will obtain the final spectrum map recovery result ${\mathcal{X}_R}$.

Image recovery refers to repairing images that partially lose local area data to restore their integrity, which is now a research hotspot in computer graphics and computer vision.
The image recovery method based on the total variation (TV) model essentially treats the image as a piecewise smooth function in bounded space.
T. Chan and J. Shen in \cite{tvrecovery} established a unified recovery model based on the principle of energy minimization, and applied it to the field of image restoration with great results. The TV model relies on the aggregation characteristics of the image when repairing, and has morphological invariance, which can recover natural images well. What's more, the TV recovery algorithm spreads the information by iteration, resulting much smoother recovery.

It is worth noting that when we perform TV-SMR operation on the 3D spectrum map, we lose the spatial correlation between the cubes in the normal direction of $XY$ plane. In order to compensate for the loss caused by this unidirectional slicing, we improve the above TV-SMR algorithm by slicing from three dimensions of $XY$, $YZ$, $ZX$, and get three sets of spectrum map slices as shown in Fig. \ref{3dtvsmr}. Then, we apply TV-SMR and obtain three ``repaired" spectrum tensors $\mathcal{X}_{RXY}$, $\mathcal{X}_{RYZ}$ and $\mathcal{X}_{RZX}$. Finally, the three recovery results are combined and averaged to get the $\mathcal{X}_R$. We call this improved algorithm as 3DTV-SMR.
Since different slicing methods capture the correlation in different directions but also lose correlation in the remaining direction, we can take advantage of this 3DTV-SMR algorithm to achieve better recovery precision.
\subsection{Complete 3D Spectrum Mapping Framework}
As discussed above, the 3D spectrum mapping framework includes pre-sampling, spectrum situation estimation, ROI-driven UAV deployment, and total variance-based spectrum map recovery. In pre-sampling, we randomly select points of ratio $r_0$. These pre-sampling points can be used as prior information to help us mine 3D spectrum situational information.
Spectrum situation estimation is the key to connect pre-sampling and UAV deployment. It effectively exploits the spectrum signal values of the sampled positions to estimate the entire 3D spectrum situation.
During the ROI-driven UAV deployment, the estimated spectrum situation and UAV flight energy consumption are both considered to select the optimal sampling locations sequentially.
When all sampling points are obtained, TV-SMR algorithm is executed to recover the undetected area.

\section{CASE STUDY: 3D SPECTRUM MAPPING}
In this section, we consider a 3D city block with a size of $100m \times 100m \times 100m$. We take a spatial granularity of $10m$ and discretize the space into a spectrum tensor of ${N_1} \times {N_2} \times {N_3} = 10 \times 10 \times 10$. We place signal sources in three positions $(0,0,0)$, $({N_1}/2,{N_2}/2,0)$, $({N_1},{N_2},0)$.
The signal power of the source is assumed to be $30mW$. The receiver noise spectral density is $ - 174dBm/Hz$ and the bandwidth is $200KHz$. Both the $\xi ( \cdot )$ and $\gamma ( \cdot )$ functions are taken as exponential functions. The UAV flying speed is $1 m /s$ and the hovering time at each sampling point is 5 seconds. The radius of each ROI is $30m$.

\begin{figure*}[!t]
\centering
\includegraphics[width=.8\linewidth]{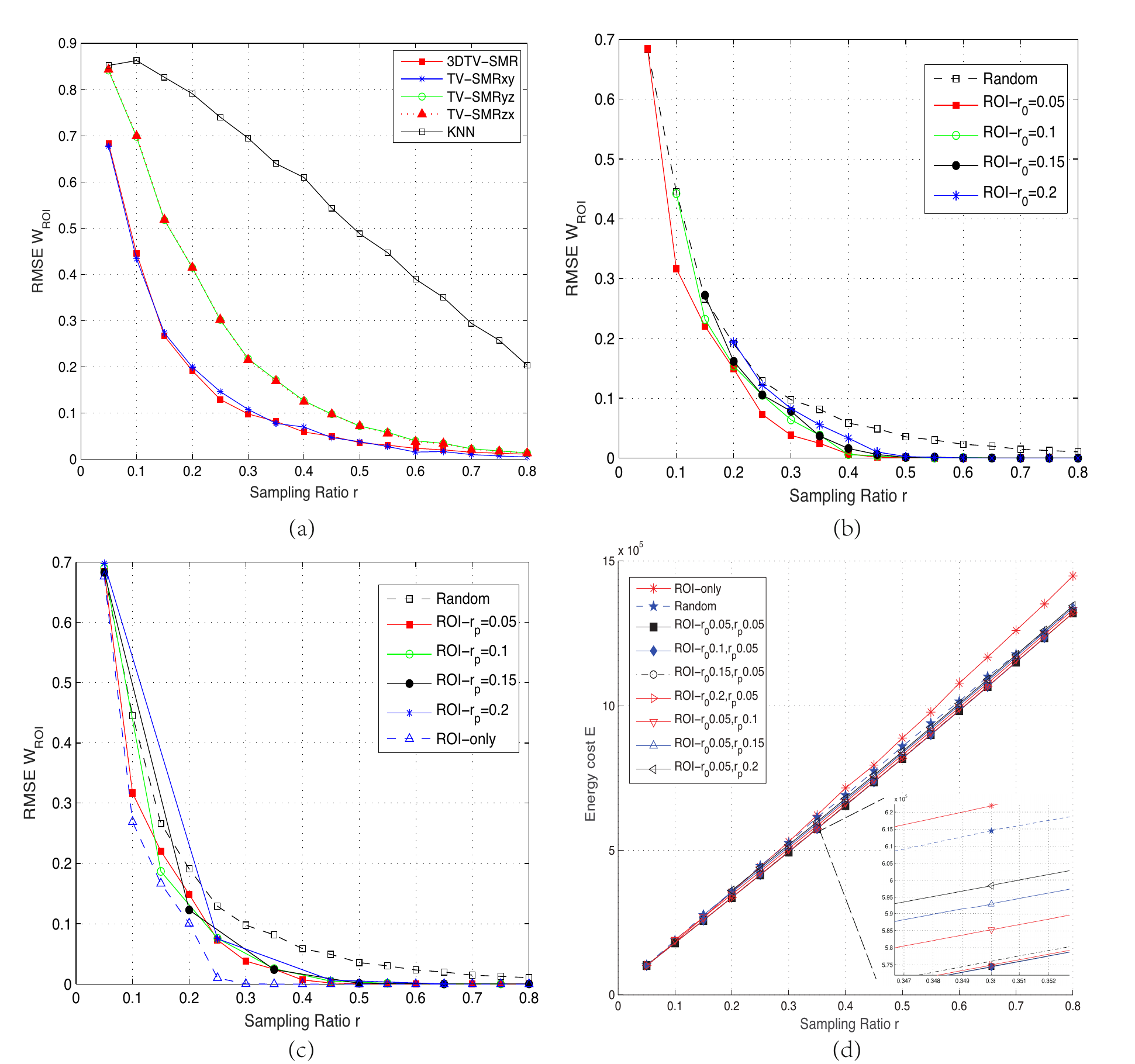}
\caption{(a) Recovery performance comparisons of 3DTV-SMR, TV-SMR and KNN with random UAV deployment.
(b) Impact of pre-sampling ratio ${r_0}$ on recovery performance comparisons of ROI-driven UAV deployment (${r_p} = 0.05$, all use 3DTV-SMR recovery algorithm). (c) Impact of newly-added samples ratio ${r_p}$ on recovery performance comparisons of ROI-driven UAV deployment (${r_0} = 0.05$, all use 3DTV-SMR recovery algorithm). (d) Energy consumption performance comparisons of ROI-only, random and ROI-driven UAV deployments as well as impact of $r_0$ and $r_p$ on energy consumption performance comparison of ROI-driven UAV deployment (all use 3DTV-SMR recovery algorithm).}
\label{simulation}
\end{figure*}

As shown in Fig. \ref{simulation} (a): $i$) For all algorithms, recovery performance decreases as the number of sampling points increases; $ii$) 3DTV-SMR and TV-SMR series algorithms are much better than the KNN algorithm on the recovery effect; $iii$) In addition, we find that the $TV-SMR_{yz}$ and $TV-SMR_{zx}$ curves coincide, and the performances of 3DTV-SMR and $TV-SMR_{xy}$ are quite close. This is because of the symmetry placement of the three sources we set, and we can further conclude that slicing in the $ZX$ and $YZ$ planes will lose more information between cubes. However, in the $XY$ plane we successfully capture the intrinsic spatial relations and features. Therefore, when we perform spectrum mapping on a completely unknown area, we can directly execute the 3DTV-SMR algorithm to avoid the trouble of finding the optimal slicing direction.

According to the Fig. \ref{simulation} (b), when the pre-sampling ratio $r_0$ increases, the performance of ROI-driven UAV deployment gradually deteriorates. Furthermore, because the pre-sampling component of our 3D spectrum mapping spectrum framework is random sampling, the start points of the four ROI curves have intersections with the random sampling curve. What's more, curves $ROI-r_0=0.1$ and $ROI-r_0=0.15$ have intersections mainly because of fluctuations caused by differences in the randomly generated sampling points, but overall $ROI-r_0=0.1$ is better than $ROI-r_0=0.15$.
For the Fig. \ref{simulation} (c), the same number of sampling points are generated in the pre-sampling component, so all ROI curves have intersections at the starting position. In addition, when the sampling rate is lower than about 0.3, the larger $r_p$ brings higher recovery accuracy. When the sampling ratio is large, the performances are similar.
In addition, it can be seen from Fig. \ref{simulation} (d) that although large $r_p$ has good performance, it will cost more energy resources.
This is because too many new sampling points are determined at one time, which leads to poor selection of points and increases UAV flying energy consumption.

In addition, compared to the ROI-driven UAV deployment and random UAV deployment schemes, here is a new deployment scheme called ROI-only deployment. In this solution, the selection principle is to select the point with the highest estimated POI value without considering energy consumption, that is,
${\cal X}_i^*{\rm{ = }}\mathop {{\rm{arg max}}}\limits_{{{\cal X}^i}} {\rm{ }}\widehat {{\cal P}{\cal O}{{\cal I}_{{{\cal X}^i}}}}$.
It can be seen from the simulation that although the performance of this scheme is better than the other two deployments, its energy consumption is the largest of the three schemes. Thus, when performance is more important than energy consumption, we can adopt this solution, otherwise 3DTV-SMR is the best choice.

\section{TECHNICAL CHALLENGES AND OPEN ISSUES}
 Here we address some of the open issues and future research directions related to 3D spectrum mapping, which still require more research efforts. These issues include:
 \begin{itemize}
\item \emph{Accurate spectrum situation estimation algorithms}. In our proposed spectrum mapping framework, spectrum situation estimation is the key between pre-sampling and UAV deployment. If the estimation error is too large, it will directly lead to poor selection of the UAV flight trajectory. However, when we combine the internal mechanism of signal transmission, namely the channel propagation characteristics, to the estimation algorithm, it will definitely improve the accuracy of the spectrum situation estimation greatly.
\item \emph{Irregular surveillance area UAV deployment}. In practical 3D environment, the size of tall buildings can't be ignored. On the one hand, this leads to restricted UAV trajectory selection. On the other hand, obstacles also affect the signal propagation in space, which brings greater challenges to the spectrum situation estimation.
\item \emph{Efficient power allocation algorithm for UAV data transmission}. When performing spectrum mapping, especially for important events, UAVs need to quickly transmit the collected spectrum data back to the data analysis center with low latency \cite{scma}. Since the power of the UAV is limited, how to allocate power for flying and data transmission with high capacity and low time delay is a huge challenge.
\item \emph{Energy consumption model for varying-speed flying UAV and speed optimization in 3D space.}
    There are multiple flying states for UAVs in 3D space, including horizontal flying, vertical flying and oblique flying. When the speed changes, the corresponding energy consumption will change non-linearly, which strongly motivates us but also makes it a great challenge to study the energy consumption model and to optimize the speed of flying UAV in 3D space.
\item \emph{Hybrid model and data driven 3D UAV spectrum recovery optimization}. We can improve the spectrum map recovery model by considering both physical signal propagation and neural network structure.
    Due to the complexity of the environment, the spectrum situation usually doesn't match the general physical model well. Thus, we can train the obtained data with the help of neural networks, which can capture the nonlinear representation of situation information. This hybrid model and data driven spectrum recovery is more conducive to selecting samples and improving the accuracy of spectrum map restoration.
 \end{itemize}

\section{CONCLUSION}
This article developed a 3D spectrum mapping based on ROI-driven UAV deployment for spectrum monitoring and management in smart IoT. We firstly presented a brief survey on the state-of-the-art studies on spectrum mapping techniques. Then, we introduced the entire system model of 3D spectrum mapping. Next, we proposed a framework for 3D spectrum mapping which is composed of pre-sampling, spectrum situation estimation, ROI-driven UAV deployment, and total variance based spectrum map recovery.
Furthermore, an exemplary simulation on the mapping framework was provided which confirms the superiority of the proposed 3D spectrum mapping framework.
Lastly, several technical challenges and open issues ahead were discussed.
We firmly believe this important area will be a fruitful research direction in smart IoT and we have just touched one tip of the iceberg. We hope this article will stimulate much more research interests.

\section{ACKNOWLEDGMENT}
This work is supported by the National Natural Science Foundation of
China (No. 61827801, No. 61801216, No. 61871398, No. 61931011, No. 61631020), the
Natural Science Foundation for Distinguished Young Scholars of Jiangsu
Province (No. BK20190030), the Natural Science Foundation of Jiangsu
Province (No. BK20180420), the open research fund of Key Laboratory
of Dynamic Cognitive System of Electromagnetic Spectrum Space (Nanjing
Univ. Aeronaut. Astronaut.), Ministry of Industry and Information Technology,
Nanjing, 211106, China (No. KF20181913).


\vspace{-40 mm}
\begin{IEEEbiographynophoto}{Qihui Wu}
[SM'13] (wuqihui2014@sina.com) received his B.S. degree in communications engineering, M.S. degree and Ph.D. degree in communications and information systems from Institute of Communications Engineering, Nanjing, China, in 1994, 1997 and 2000, respectively. From 2003 to 2005, he was a Postdoctoral Research Associate at Southeast University, Nanjing, China. From 2005 to 2007, he was an Associate Professor with the College of Communications Engineering, PLA University of Science and Technology, Nanjing, China, where he served as a Full Professor from 2008 to 2016. Since May 2016, he has been a full professor with the College of Electronic and Information Engineering, Nanjing University of Aeronautics and Astronautics, Nanjing, China. From March 2011 to September 2011, he was an Advanced Visiting Scholar in Stevens Institute of Technology, Hoboken, USA. Dr. Wu's current research interests span the areas of wireless communications and statistical signal processing, with emphasis on system design of software defined radio, cognitive radio, and smart radio.
\end{IEEEbiographynophoto}

\vspace{-40 mm}
\begin{IEEEbiographynophoto}{Feng Shen}
(sfjx$\_$nuaa@163.com) received his B.S. degree in information engineering from Nanjing University of Aeronautics and Astronautics, Nanjing, China, in 2017. He is currently pursuing his Ph.D degree at the College of Electronics and Information Engineering, Nanjing University of Aeronautics and Astronautics, Nanjing, China. His research interests include cognitive information theory, cognitive radio, signal processing and wireless communications.
\end{IEEEbiographynophoto}

\vspace{-40 mm}
\begin{IEEEbiographynophoto}{Zheng Wang}
(z.wang@ieee.org) received the B.S. degree in electronic and information engineering from Nanjing University of Aeronautics and Astronautics, Nanjing, China, in 2009, and the M.S. degree in communications from the Department of Electrical and Electronic Engineering, University of Manchester, Manchester, U.K., in 2010. He received the Ph.D degree in communication engineering from Imperial College London, UK, in 2015.
From 2015 to 2016 he served as a Research Associate at Imperial College London, UK and from 2016 to 2017 he was an senior engineer with Radio Access Network R$\& $D division, Huawei Technologies Co.. He is currently an assistant professor at the College of Electronics and Information Engineering, Nanjing University of Aeronautics and Astronautics (NUAA), Nanjing, China. His current research interests include lattice methods for wireless communications, cognitive radio and physical layer security.
\end{IEEEbiographynophoto}

\vspace{-40 mm}
\begin{IEEEbiographynophoto}{Guoru Ding}
(dr.guoru.ding@ieee.org) is an associate professor in the College of Communications Engineering, Nanjing, China. His research
interests include cognitive radio networks, massive MIMO, machine learning, and big data analytics over wireless networks. He is now an associate editor of IEEE Transactions on Cognitive Communications and Networking. He served as a guest editor for the IEEE Journal on Selected Areas in Communications (special issue on spectrum sharing and aggregation in future wireless networks).

\end{IEEEbiographynophoto}


\begin{thebibliography}{99}

\bibitem{IoT-survey}
N. H. Motlagh, T. Taleb, and O. Arouk, ``Low-altitude unmanned aerial vehicles-based internet of things services: Comprehensive survey and future perspectives,'' \emph{IEEE Internet of Things Journal}, vol. 3, no. 6, pp. 899-922, Dec. 2016.

\bibitem{fcc}
FCC, ``Facilitating opportunities for flexible, efficient, and reliable spectrum use employing cognitive radio technologies,'' \emph{ET Docket 03-108}, Dec. 2003.

\bibitem{radiomap}
J. Schuette, B. Fell, J. Chapin, S. Jones, J. Stutler, M. Birchler, and D. Roberson, ``Performance of RF mapping using opportunistic distributed devices," in \emph{2015 IEEE Military Communications Conference}, Tampa, FL, USA, Oct. 2015, pp. 1624-1629.

\bibitem{futureiot}
Q. Wu, G. Ding, Z. Du, Y. Sun, M. Jo, and A. V. Vasilakos, ``A cloud-based architecture for the internet of spectrum devices over future wireless networks,'' \emph{IEEE Access}, vol. 4, pp. 2854-2862, Jun. 2016.

\bibitem{uavpopularity}
G. Ding, Q. Wu, L. Zhang, Y. Lin, T. A. Tsiftsis, and Y.-D. Yao, ``An amateur drone surveillance system based on the cognitive internet of things,'' \emph{IEEE Communications Magazine}, vol. 56, no. 1, pp. 29-35, Jan. 2018.

\bibitem{PrioritizedSensing}
E. Ate{\c{s}}, T. E. Kalayci, and A. U{\u{g}}ur, ``Area-priority-based sensor deployment optimisation with priority estimation using K-means,'' \emph{IET Communications}, vol. 11, no. 7, pp. 1082-1090, May 2017.

\bibitem{rem}
V. Atanasovski, J. van de Beek, A. Dejonghe, D. Denkovski, L. Gavrilovska, S. Grimoud, P. M{\"a}h{\"o}nen, M. Pavloski, V. Rakovic, J. Riihijarvi et al., ``Constructing radio environment maps with heterogeneous spectrum sensors,'' in \emph{2011 IEEE International Symposium on Dynamic Spectrum Access Networks (DySPAN)}, Aachen, Germany, May 2011, pp. 660-661.

\bibitem{Dingkernel2013}
G. Ding, Q. Wu, Y. Yao, J. Wang, and Y. Chen, ``Kernel-based learning for statistical signal processing in cognitive radio networks: Theoretical foundations, example applications, and future directions,'' \emph{IEEE Signal Processing Magazine}, vol. 30, no. 4, pp. 126-136, Jul. 2013.

\bibitem{uav3dsensing2019}
F. Shen, G. Ding, Z. Wang, and Q. Wu, ``UAV-based 3D spectrum sensing in spectrum-heterogeneous networks,'' \emph{IEEE Transactions on Vehicular Technology}, vol. 68, no. 6, pp. 5711-5722, Jun. 2019.

\bibitem{sun2018}
J. Sun, J. Wang, G. Ding, L. Shen, J. Yang, Q. Wu, and L. Yu, ``Long-term spectrum state prediction: An image inference perspective,'' \emph{IEEE Access}, vol. 6, pp. 43489-43498, Jul. 2018.

\bibitem{tsp}
G. Reinelt, \emph{The Traveling Salesman: Computational Solutions for TSP Applications}. Springer-Verlag, 1994.

\bibitem{idw}
O. Huisman and R. De By, ``Principles of geographic information systems,'' \emph{ITC Educational Textbook Series}, vol. 1, pp. 334-352, 2009.

\bibitem{energymodel}
H. V. Abeywickrama, B. A. Jayawickrama, Y. He, and E. Dutkiewicz, ``Comprehensive energy consumption model for unmanned aerial vehicles, Based on empirical studies of battery performance,'' \emph{IEEE Access}, vol. 6, pp. 58383-58394, Oct. 2018.

\bibitem{tvrecovery}
J. Shen and T. F. Chan, ``Mathematical models for local nontexture inpaintings,'' \emph{SIAM Journal on Applied Mathematics}, vol. 62, no. 3, pp. 1019-1043, Feb. 2002.

\bibitem{scma}
S. Han, Y. Huang, W. Meng, C. Li, N. Xu, and D. Chen, ``Optimal power allocation for SCMA downlink systems based on maximum capacity,'' \emph{IEEE Transactions on Communications}, vol. 67, no. 2, pp. 1480-1489, Feb. 2019.

\end{thebibliography}
\end{document}